\RequirePackage{fixltx2e}
\documentclass[aps,pre,preprint,showpacs,showkeys,superscriptaddress,nofootinbib,longbibliography,floatfix,a4paper]{revtex4-1}
\usepackage{amsmath,amssymb,graphicx,bbm,enumerate,theorem}
\usepackage[caption=false]{subfig}
\usepackage[english]{babel}
\usepackage{xcolor}
\usepackage{dcolumn}
\usepackage{mathtools}
\usepackage[section]{placeins}

\newcommand{\mathe}{\mathrm{e}}
\newcommand{\tmop}[1]{\ensuremath{\operatorname{#1}}}

\begin{document}

\title{Fast computation method for comprehensive agent-level epidemic
  dissemination in networks} 
%\date{\today}
\author{Gilberto M. Nakamura}
%\email{gmnakamura@usp.br}
 \affiliation{Universidade de S\~{a}o Paulo, Ribeir\~{a}o Preto 14040-901, Brazil}
 \author{Ana Carolina P. Monteiro}
% \email{ana.carolina.monteiro@usp.br}
 \affiliation{Universidade de S\~{a}o Paulo, Ribeir\~{a}o Preto 14040-901, Brazil}
 \author{George C. Cardoso}
\email{gcc@usp.br}
 \affiliation{Universidade de S\~{a}o Paulo, Ribeir\~{a}o Preto  14040-901, Brazil}
\author{Alexandre S. Martinez}
%\email{asmartinez@ffclrp.usp.br}
 \affiliation{Universidade de S\~{a}o Paulo, Ribeir\~{a}o Preto 14040-901, Brazil}
\altaffiliation{Instituto Nacional de Ci\^{e}ncia e Tecnologia - Sistemas Complexos (INCT-SC)}

\begin{abstract}
Two simple agent based models are often employed in epidemic studies: the
susceptible-infected (SI) and the susceptible-infected-susceptible
(SIS). Both models describe the time  evolution of infectious diseases
in networks in which vertices are either susceptible (S) or infected (I) agents. 
Predicting the effects of disease spreading is one of the major goals in
epidemic studies, but often restricted to numerical simulations. Analytical
methods using operatorial content are subjected to  the asymmetric
eigenvalue problem, restraining the usability of standard perturbative
techniques, whereas numerical methods are limited to small populations since
the vector space increases exponentially  with population size $N$.
Here, we propose the use of the squared norm of probability vector, $\vert P(t)\vert
^2$, to obtain an algebraic equation which allows the evaluation of
stationary states, in time independent Markov processes.
The equation requires eigenvalues of symmetrized time generators, which 
take full advantage of system symmetries, reducing 
the problem to an $O(N)$ sparse matrix diagonalization.
 Standard perturbative methods are
introduced, creating precise tools to evaluate the effects of health
policies. 
\end{abstract}
%02.50.Ga	Markov processes
%05.10.Gg	Stochastic analysis methods
%64.60.aq	Networks
%89.75-k        Complex systems

\maketitle

One of the main goals in epidemics studies of communicable diseases is
to correctly predict the time evolution of a given disease within a
 population \cite{martinezJStatPhys2011}. The forecasting procedure,
 which may take numerical or  analytic formulations, often 
encounters obstacles due to heterogeneous populations and the disease
spreading dynamics. For instance, ambiguous symptoms among distinct
diseases may under or overestimate total reported infections, leading
to incorrect estimates of transmission rates. 
% exhibits a set of particularities, ranging from symptoms and
% transmission mechanisms to synergies with other diseases. 
Several epidemic models have been tailored to better grasp general
behaviors in disease spreading
\cite{kermack1991,satorrasRevModPhys2015}. Among them, the simplest one
is the 
susceptible-infected-susceptible (SIS) model. The SIS model is a
Markov process and describes the time evolution of a single infectious
disease in a population formed by susceptible (S) and infected
agents (I). The infected agents carry the disease pathogens and may
transmit them to susceptible agents with constant transmission rate 
$\beta$. The model also contemplates cure events for infected agents
with constant cure rate $\gamma$ and so does competition
between cure and infection events.

There are two popular approaches often employed to mimic the disease
spreading dynamic in populations with fixed size $N$: compartmental
and stochastic ones \cite{heesterbeekScience2015}. In the compartmental approach, relevant
properties derived from either infected or susceptible agents are
well-described by averages, a direct result from the
random-mixing hypothesis \cite{bansalJRSoc2007}. This enables one to derive
non-linear differential equations to match the evolution of disease
throughout the population. For instance, the number of infected agents
in the compartmental SIS model, $n(t)$, satisfies the following
differential equation: 
\begin{equation}
  \frac{d n}{dt} = \frac{\beta}{N}\frac{\langle k\rangle}{N} n(N-n)-\gamma n,
\label{eqcompartimental1}
\end{equation}
with $\langle k\rangle = N-1$ and basic reproduction number
\cite{feffernanJRSoc2005} $R_0=\beta/\gamma$ . For homogeneous  
populations, this is the  
expected behavior. However, real agents differ from each other, leading
to heterogeneous population, in disagreement with the random-mixing
hypothesis \cite{keelingPNAS2002}.   
% A single realization of a given network is represented by one graph,
% which is a mathematical structure labeled by a set of nodes and a set of
% link among the nodes. The link distribution follows the network
% distribution rule .
Stochastic approaches may also be further classified according to
their descriptive variable. Similar to the compartmental model, the
mesoscopic interpretation usually describes the time evolution of global
variables
\cite{caliriPhysicaA2003,caliriJBioPhys1999},
however, it allows fluctuations along time. Meanwhile, the microscopic
approach describes the disease spreading of individual agents and their
interactions, thus introducing fluctuations at the agent level over
time.  Both approaches mostly differ on how they treat fluctuations due
to agent heterogeneity within a given population, in the epidemic
processes.

Central to the microscopic stochastic approach is the underlying
network used to reproduce the heterogeneity typically found within
populations \cite{keelingJRSoc2005}. In the network scheme, agents
are represented by vertices and their connections are distributed
according to the adjacency matrix $A$ for the assigned network
configuration (graph) \cite{albertRevModPhys2002}.   
In this case, it is well-accepted that the mean number $\langle
k\rangle$ of vertex connection in Eq.~(\ref{eqcompartimental1}) describes the
averaged process. Contrary to the random mixing hypothesis,
non-trivial topological aspects of $A$ may be incorporated in the
effective transmission and cure rates, producing complex patterns in
epidemics \cite{satorrasRevModPhys2015}.  
The time evolution is dictated by the transition matrix $T$, whose
matrix elements $T_{\mu \nu}$ are transition probabilities from network
configuration $\nu$ to $\mu$ \cite{vanKampen1981}. In general, one often
assumes Markovian behavior to describe disease transmission and cure
events, in accordance with the Poissonian assumption
\cite{satorrasRevModPhys2015}.   
% Now, both compartmental and stochastic formulations greatly differ on
% the way they approach fluctuations and their role in epidemics. In the
% compartmental version, fluctuations are byproducts of topological
% effects. In fact, for a single network realization or even regular
% graphs, compartmental dynamics produces no fluctuations whereas 
% fluctuations are intrinsic components in the stochastic formulation
% \cite{vanKampen1981}. 
The difference between the compartmental and stochastic schemes leads to
distinct evolution patterns for statistics as well. For instance,
Eq.~(\ref{eqcompartimental1}) displays stable infected population for
$\gamma<\beta$, power-law behavior for $\gamma=\beta$ and exponential
decay otherwise. While all three behaviors are also observed in the
stochastic approaches, fluctuations become much more relevant when the
number of infected agents, $\langle n(t)\rangle$, is small compared to 
total population, $N$. Incidentally, this is the relevant regime
to sanitary measures and health policies to contain real epidemics in
early stages.

%  which is important for prevention. This
% ultimately brings the power-law behavior 
%  for $\gamma/\beta < 1$. For instance, in the fully connected network, $A_{i
%    j}=(1-\delta_{i j})$, stable infected population holds for
%  $\gamma/\beta < 3/4$.

The Markovian approach produces accurate results if the infection
transmission is known. However, its usability is restricted to
numerical simulations with small $N$ since computational time is $O(N^2)$. This weakness lies in the
fact the $\hat{T}$ is generally non-hermitian
\cite{vanKampen1981}. Therefore, left and 
right eigenvectors are not related by transpositions, limiting the
exact diagonalization only to small values of $N$ or special
transition matrices. 
One of the main goals in epidemic studies is the ability to correctly
predict how small parameter or topological changes in the network
affect the disease spreading. If such predictions are robust,
preemptive actions to lessen the epidemic are also expected to achieve
better results. This is exactly the subject of perturbation
techniques, which make extensive use of scalar product between left and
right eigenvectors. In epidemic models, however, one must deal with
asymmetric transitions, prohibiting perturbative schemes based on
normed scalar products.

Here, we have devised a method to avoid difficulties related to the
non-hermiticity of $T$ using the squared norm of probability vector,
$\lvert P(t)\rvert ^ 2$. 
The proposed method allows us to obtain one differential equation, which
relies only on eigenvalues of symmetrized time generators. In
particular, the equation resumes to an algebraic equation for
stationary states. More importantly, the proposed method allows for
seamless reproduction of
traditional perturbative results, shedding light on the
role played by fluctuation in epidemics.
In Sec. \ref{sec:conf}, we introduce the mathematical aspects regarding
the agent based SIS model, with emphasis on the transition matrix
and operatorial content. In Sec. \ref{sec:time_evol}, we discuss calculation of
statistics and the temporal equation for $\lvert P(t)\rvert ^2$. Results
obtained using perturbative methods for epidemics in agent based
networks for SIS model are shown in
Sec. \ref{sec:perturbation}. Finally, the main conclusions are stated in
Sec. \ref{sec:conclusion}.

 \section{Transition matrix}
 \label{sec:conf}

\begin{figure}[ht]
  \begin{tabular}{ccc}
    \includegraphics[width=0.4\columnwidth]{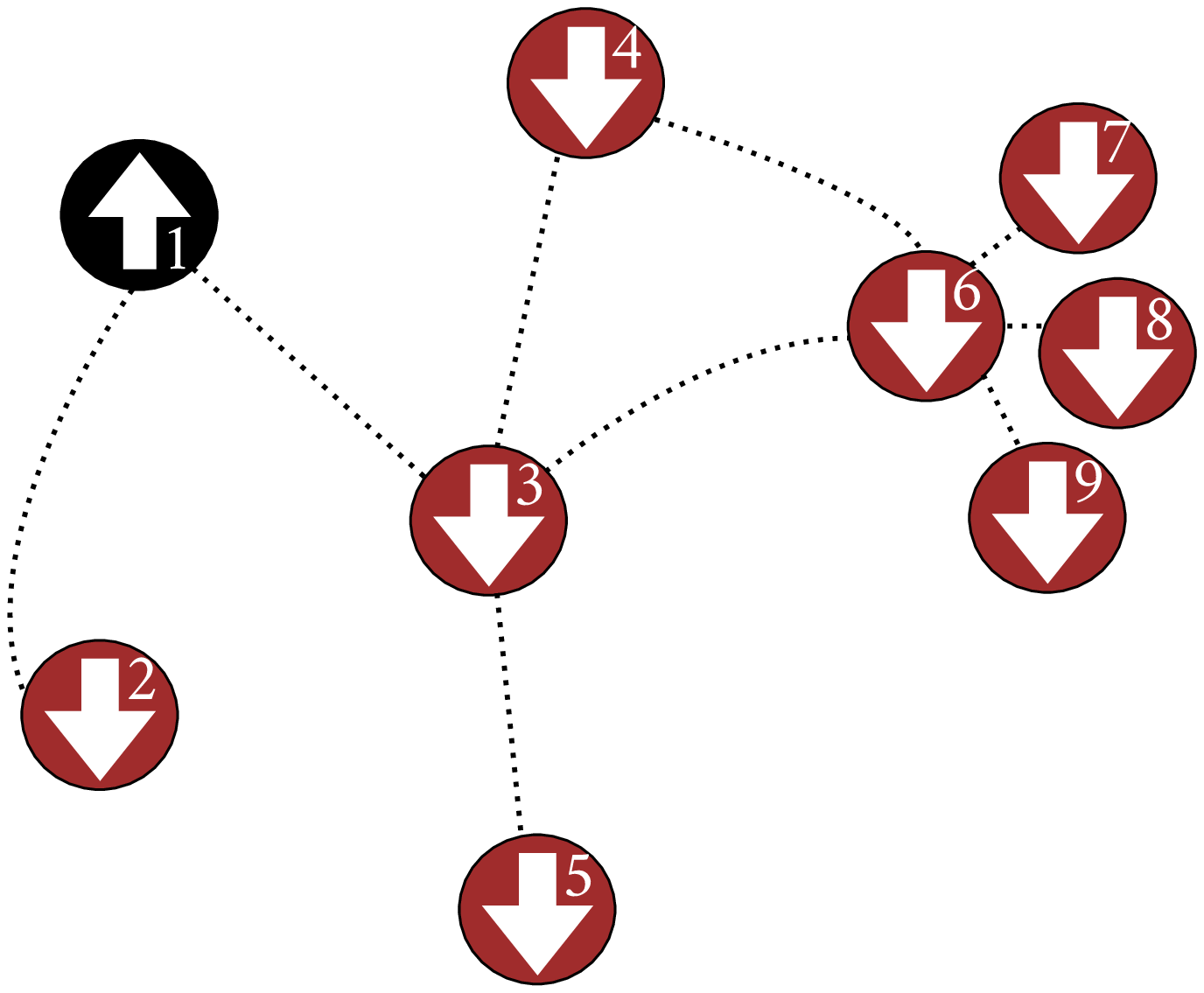} &
    &\includegraphics[width=0.4\columnwidth]{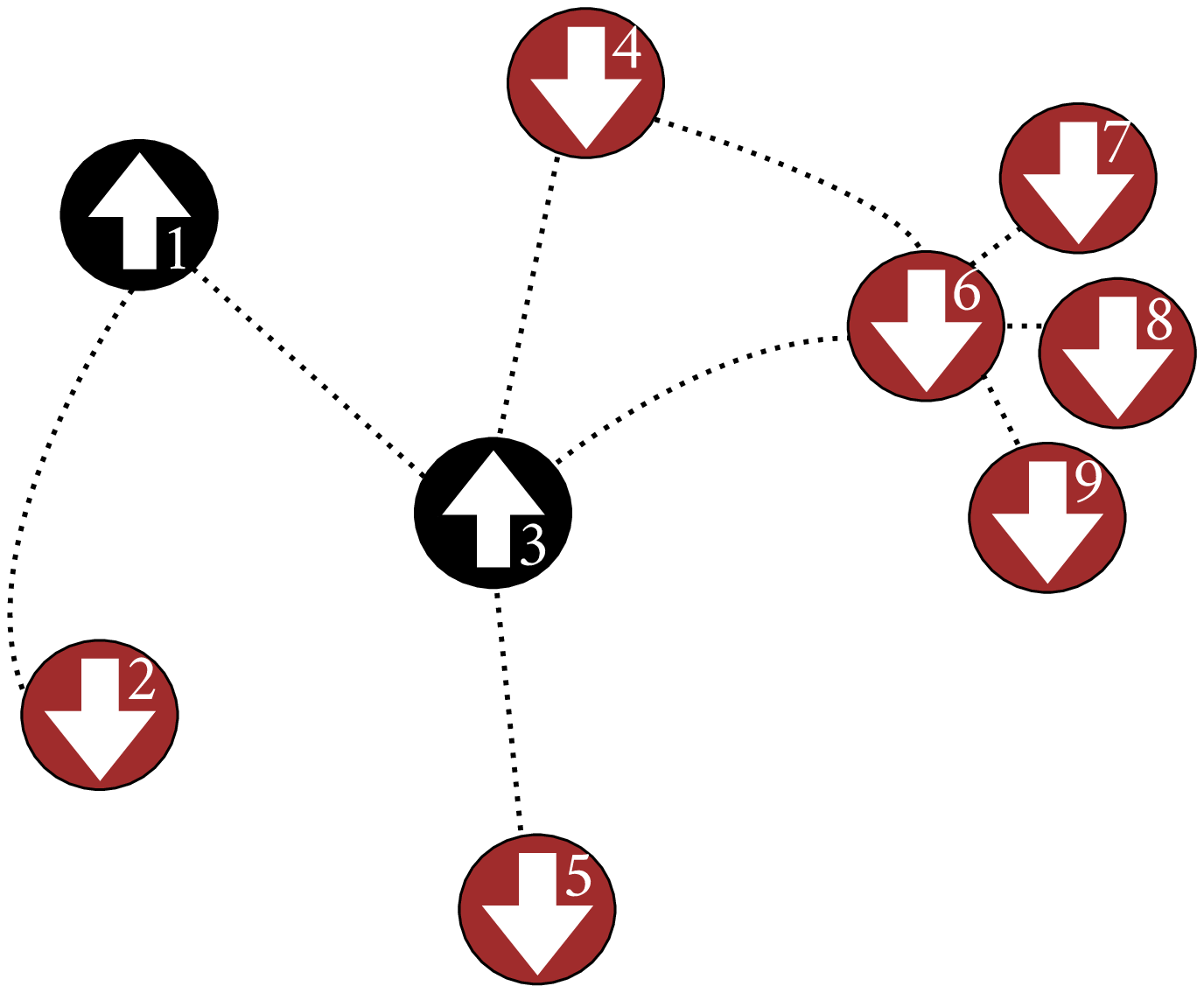}\\
    a) & &b) 
  \end{tabular}
  \caption{ \label{fig:conf_example} 
    Agent configurations in epidemic process. In a), agent at vertex
    $k=1$ is infected. The graph configuration is $ C_{1}=\lvert
    \uparrow \downarrow \downarrow \downarrow   \downarrow
    \downarrow \downarrow \downarrow \downarrow \rangle$. In
    b), a second agent is infected at $k = 3$, leading to the  configuration  
    $C_5= \lvert \uparrow \downarrow \uparrow \downarrow \downarrow
    \downarrow \downarrow \downarrow \downarrow \rangle$.}
\end{figure}

Graphs are mathematical realizations of networks
\cite{albertRevModPhys2002}. They are formed by a  set of interconnected
vertices $V_k$ ($k = 1, \ldots, N$). The
connections are described by the adjacency matrix $A$ ($N\times N$),
whose matrix elements are either  $0$ or $1$. Vertex $k$ is connected
to vertex $k'$ if $A_{k k^{\prime}}=1$, and $0$ otherwise. Within our framework, each
vertex $V_k$  holds a single agent, whose current status is identified
by $\sigma_k$. The variable $\sigma_k$ may acquire two values, namely,
$\sigma_k = \downarrow$ (susceptible) or $\sigma_k = \uparrow$
(infected), fulfilling the two-state requirement. The configuration
$C_{\mu}$ describes all agent states in the graph, $\lvert C_{\mu}
\rangle \equiv \lvert \sigma_1 \sigma_2 \cdots \sigma_N \rangle$, 
with $\mu = 0, 1, \ldots, 2^N - 1$, as shown in Fig.~\ref{fig:conf_example}. For lack of a better procedure, we 
enumerate the configurations using binary arithmetic: $\mu =
\delta_{\sigma_1 \uparrow}
2^0 + \delta_{\sigma_2 \uparrow} 2^1 + \cdots + \delta_{\sigma_N \uparrow}
2^{N - 1}$, where the Kronecker delta $\delta_{\sigma_k\uparrow}=1$ if
$\sigma_k=\uparrow$ and null otherwise. The set $\{ C_{\mu} 
\}$  spans a discrete Hilbert space $\mathbbm{H}$. For clarity's sake,
we use the following notation: Latin indices  run over vertices
$1, \ldots, N$, while Greek indices enumerate $2^N$ configurations in 
$\mathbbm{H}$. 

The next step required to assemble the transition matrix is the
definition of operators and their actions on vectors in $\mathbbm{H}$.
For instance, the operator $\hat{\sigma}^z_k$ probes whether the agent located at
vertex $k$ is infected ($\uparrow$) or not ($\downarrow$):
\begin{equation}
  \hat{\sigma}^z_k | \sigma_1 \sigma_2 \cdots \sigma_N \rangle = (
  \delta_{\sigma_k \uparrow} - \delta_{\sigma_k \downarrow}) | \sigma_1
  \sigma_2 \cdots \sigma_N \rangle,
\end{equation}
while the operator $\hat{n}_k={(\hat{\sigma}^z_k + 1)}/{2}$ extracts the
number of infected agent at vertex $k$.
Accordingly, the operator $\hat{n} = \sum_k \hat{n}_k$ extracts the
total number of infected agents in the population. The $k$-th agent health
status is switched by action of operators $\hat{\sigma}_k^+$ and $\hat{\sigma}_k^-$:
\begin{align}
  \hat{\sigma}_k^+ \lvert\sigma_1\cdots \downarrow\cdots\sigma_N
  \rangle =& \lvert \sigma_1\cdots \uparrow \cdots\sigma_N\rangle, \\
  \hat{\sigma}_k^- \lvert \sigma_1\cdots \uparrow
  \cdots\sigma_N\rangle =& \lvert \sigma_1\cdots \downarrow
                           \cdots\sigma_N\rangle, 
\end{align}
null otherwise. Another useful operator is
$\hat{\sigma}_k^x=\hat{\sigma}_k^++\hat{\sigma}_k^-$. 
The localized operators $\hat{\sigma}_k^{\pm}$ and $\hat{\sigma}_k^z$
satisfy additional algebraic properties. For each $k$, the set
$\hat{\sigma}_k^{\pm, z}$ forms a local $\tmop{su} ( 2)$ algebra,
$  [ \hat{\sigma}_k^z, \hat{\sigma}_k^{\pm}]  =  \pm 2
  \hat{\sigma}_k^{\pm}$, $ [\hat{\sigma}_k^+, \hat{\sigma}_k^-]  =
  \sigma_k^z$ and 
$\{ \hat{\sigma}_k^+, \hat{\sigma}_k^- \}  =  1$.
Note that $\sigma$ operators satisfy local
fermionic anticommutation relations \cite{liebmattis}. However,
their non-local algebraic commutation relations are bosonic:  
$[\hat{\sigma}_k^{\beta}, \hat{\sigma}_{k'}^{\gamma}] = 0$, for $k\neq k'$ 
and $\beta, \gamma = \pm, z$.

%------------------------------------------------------------

Let $P_{\mu}(t)$ be the probability to find the system in
the configuration $\lvert C_{\mu}\rangle$, at time $t$. The collection
of all $P_{\mu}(t)$ forms the probability vector, $\lvert P(t)\rangle
=\sum_{\mu} P_{\mu}(t)\lvert C_{\mu}\rangle$, with
$\sum_{\mu}P_{\mu}(t)=1$.
For any Markov process, the transition matrix $\hat{T}$ describes allowed
transitions among configurations such that $\lvert P(t+\delta t)\rangle =
\hat{T}\lvert P(t)\rangle$. 
Under Poissonian assumption \cite{satorrasRevModPhys2015}, one
only considers either a single cure or single infection event during a time
interval $\delta t$. The Poissonian hypothesis tends to be more accurate
for vanishing $\delta t$.

\begin{figure}[htb]
  \includegraphics[width=0.8\columnwidth]{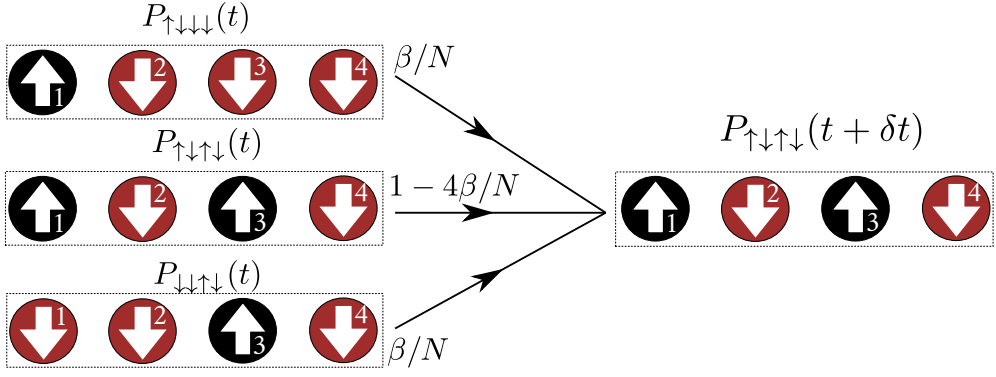}
  \caption{Markov process during the transmission phase. The probability
    $P_{\uparrow \downarrow 
      \uparrow \downarrow} ( t + \delta t)$ to find the system in the
    configuration $\lvert  \uparrow \downarrow \uparrow \downarrow
    \rangle$, at time $t + \delta t$, depends on probability
    $P_{\uparrow \downarrow \downarrow \downarrow} (t)$ that the system was 
    previously in the configuration
    $\lvert\uparrow\downarrow\downarrow\downarrow\rangle$  and then  
    transitioned with conditional probability $p (\uparrow \downarrow \uparrow
    \downarrow | \uparrow \downarrow \downarrow \downarrow) = \beta / N$ to
    state
    $\lvert\uparrow\downarrow\uparrow\downarrow\rangle$. Analogous
    rationale applies to the configuration $\lvert 
    \downarrow\downarrow\uparrow\downarrow \rangle$.  The other  
    possibility  is that the system was already in the state $\lvert\uparrow
    \downarrow \uparrow \downarrow \rangle$ at time $t$ and remains unchanged
    during the time interval $\delta t$. As such, the probability to
    remain unchanged equals to one minus the probability to change to any other
    state. In this example, the graph is fully connected and there are
    $4$ such transitions.  \label{fig:infection_example}} 
\end{figure}

\begin{figure}
  \includegraphics[width=0.80\columnwidth]{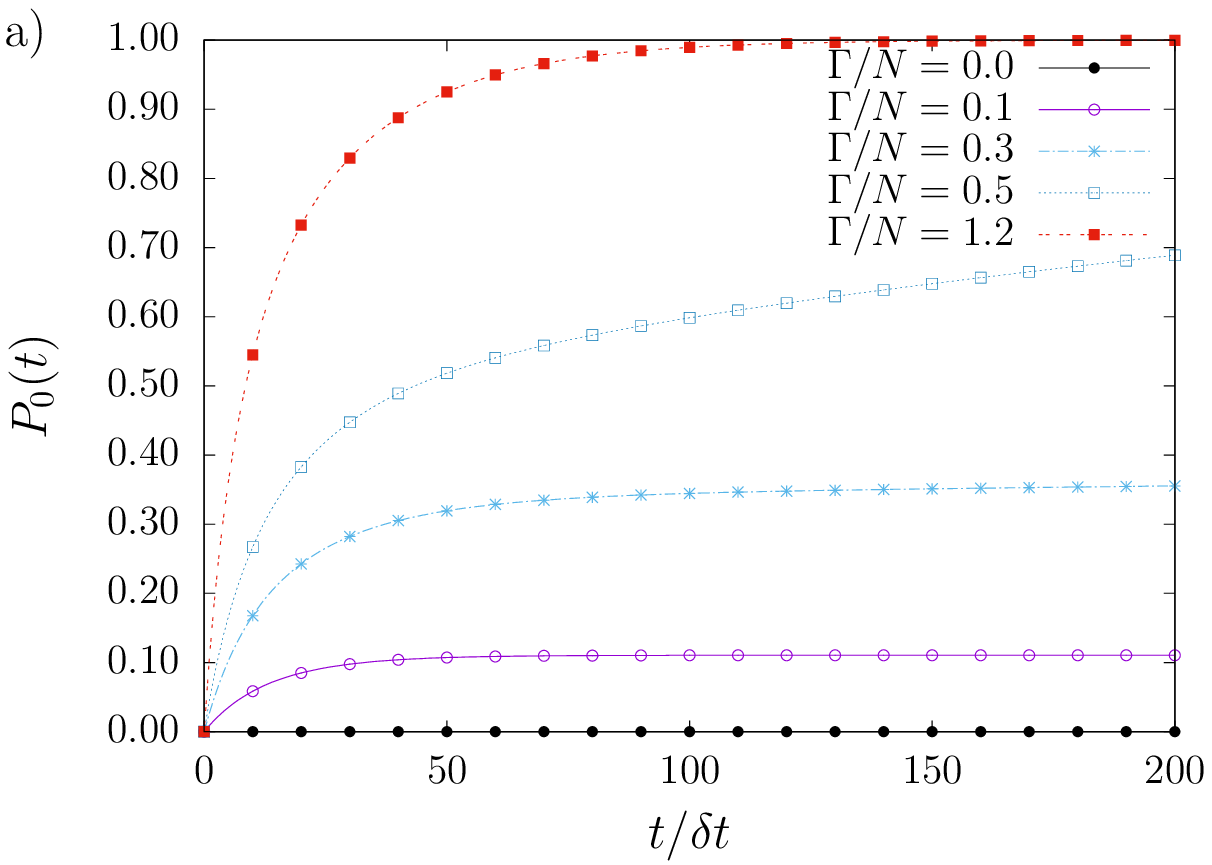}\\  
  \includegraphics[width=0.470\columnwidth]{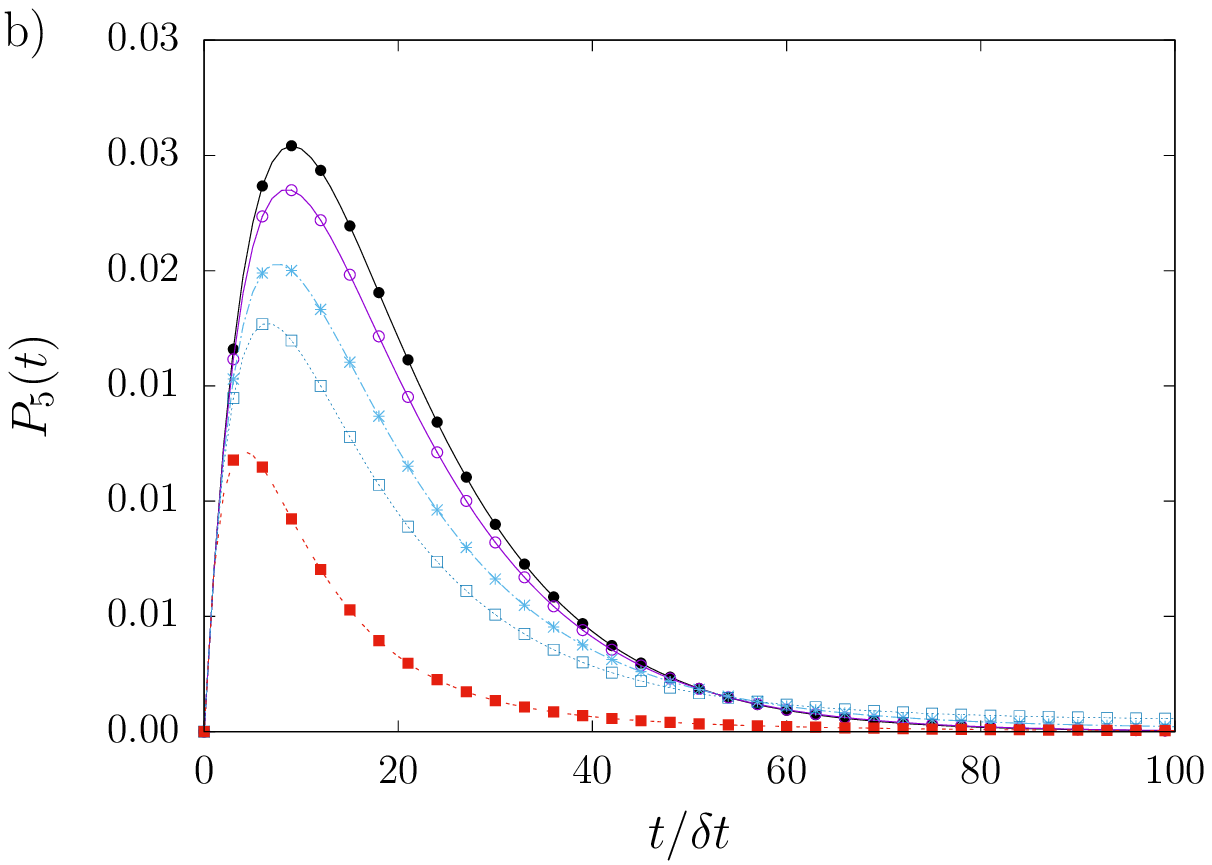} 
  ~
  \includegraphics[width=0.470\columnwidth]{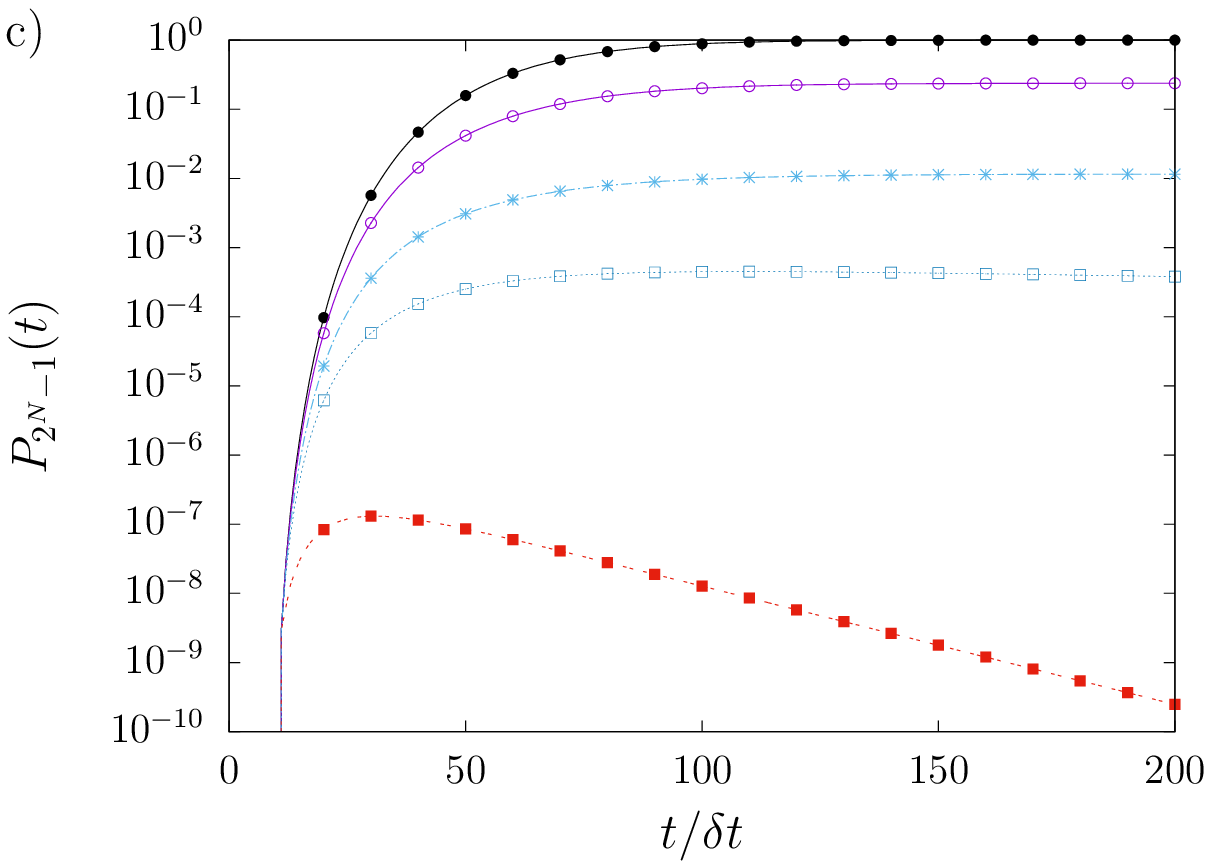}
  % \begin{tabular}{ccc}
  %   \includegraphics[width=0.21\columnwidth]{figure3b.eps} 
  %   &
  %   &
  %   \includegraphics[width=0.21\columnwidth]{figure3c.eps}
  % \end{tabular}
  \caption{\label{fig:conf_probs} Configuration probabilities
    $P_{\mu}(t)$ for $\mu=0,5$ and $2^N-1$ with $N=12$ in a fully
    connected network. In a), probability $P_0(t)$
    to observe all-cured configuration at time $t$ for various couplings
    $\Gamma/N$. In b), $P_5(t)$ refers to the probability of transient 
    configuration $\lvert
    \uparrow_1\uparrow_2\downarrow_3\cdots\downarrow_N\rangle$, while c)
    exhibits the probability with all-infected agents, $P_{2^N-1}(t)$,
    in log scale. The coupling 
    $\Gamma/N=0$ represents the SI model, whose stationary state is
    described by all-infected configuration. For $\Gamma/N=0.1$, the
    stationary state is a linear combination of distinct
    $C_{\mu}$, including all-cured $C_0$ and all-infected $C_{2^N-1}$
    configurations. For intermediate couplings $\Gamma/N=0.3$ and 
    $\Gamma/N=0.5$, the stationary state is $C_0$ with large transient
    $\Delta \tau\sim o(N^4)$. }
\end{figure}

In the SIS model, any previously infected agent at vertex $k$ is
subjected to three distinct outcomes during the time interval $\delta
t$: transmit the disease to one connected susceptible agent; cure
itself; or remain unchanged. The operator $\hat{\sigma}^{-}_k\hat{n}_k$
produces the desired cure action, while $A_{k
  m}\hat{\sigma}^{+}_m \hat{n}_k$ transmits the disease from the $k$-th
agent to $m$-th agent, given the $k$-th agent is currently infected and
the other is susceptible,
as exemplified for the fully connected graph depicted in Fig.~\ref{fig:infection_example}. 
If the cure and infection phases are independent from each other then
$\hat{T}=\hat{T}_{\text{cure}} \hat{T}_{\text{infec}}$.  Under this 
circumstances, the transition matrix is
\begin{equation}
  \hat{T}=\mathbbm{1}- \frac{\beta}{N}\sum_{k j} \left[   A_{j k} ( 1
    -\hat{n}_j - \hat{\sigma}^{+}_j) + \Gamma \delta_{k
      j}(1-\hat{\sigma}^{-}_j) \right] \hat{n}_k ,
\label{eqtsis}
\end{equation}
with $\Gamma=\gamma N/ \beta$. Once the explicit action of $\hat{T}$ 
is known, $P_{\mu}(t)$ are readily evaluated. 
Fig.~\ref{fig:conf_probs} exhibits numerical results for $P_{\mu}(t)$ for
$\mu=0,5, 2^N-1$, parameter $\Gamma/N = 0.0, 0.1, 0.3, 0.5,
1.2$, $P_1(0)=1$ as initial condition and $N=12$, in a fully connected network. 
For increasing $\Gamma/N$, the probability $P_0(t)$ to find the system
without infected agents also increases, while the opposite holds true
for $P_{2^N-1}(t)$, in which all agents are infected. The intermediate
configuration $C_5$ is displayed to emphasize transient effects. Despite its
simplicity, Eq.~(\ref{eqtsis}) produces power-law behavior,
exemplified in Fig.~\ref{fig:conf_probs} with $\Gamma/N=0.3, 0.5$, in
which the time interval to reach the stationary state $C_0$ is much
larger than the total number of agents,  $\Delta \tau\gg N\delta t$. 
Brief inspection reveals $\hat{T}$ is
asymmetric, thus implying the existence of distinct left and right 
eigenvectors.

% One single network realization is considered to derive
% Eq.~(\ref{eqtsis}). 
Derivation of Eq.~(\ref{eqtsis}) considers only a single network
realization. If an ensemble containing $M$ graphs is considered,
properly sampling the network, the only modification required is the
following: $A_{j k}\rightarrow \bar{A}_{j k} = M^{-1}\sum_{l=1}^M A_{j
  k}^{(l)}$.  The reason is the following: networks only assign
distribution rules for connections,
% since the connection
% distribution within a network only affects the connections, 
leaving the vertex distribution and, therefore, the Hilbert space
unchanged. For each graph $l=1,\ldots, M$ in the ensemble, one applies
the associated transition matrix, $\hat{T}^{(l)}$, on the initial
configuration $\lvert  P^{(l)}(0)\rangle$, producing the probability
vector $\lvert P^{(l)}(\delta t)\rangle$. In this way, one must also
consider the ensemble averages. In particular, the average probability
to find the system in configuration $\lvert C_{\mu} \rangle$ is
%\begin{equation}
$\langle P_{\mu}\rangle_{M} = {M}^{-1}\sum_{l=1}^M {P_{\mu}^{(l)}(t)}$.
%\end{equation}
Since the procedure is equivalent to the average of $\hat{T}$ over
the graph ensemble -- the network sample -- one needs only to consider
the network distribution of $A$. For clarity, we drop the bar symbol
and always assume the average over graph ensemble.

%--------------------------------------------------------

\section{Squared Norm}
\label{sec:time_evol}

\begin{figure}[htp!]
%  \begin{tabular}{lc}
%    a)&\includegraphics[width=0.45\columnwidth]{figure4a.eps}\\
%    b)&\includegraphics[width=0.45\columnwidth]{figure4b.eps}
%  \end{tabular}
  \includegraphics[width=0.80\columnwidth]{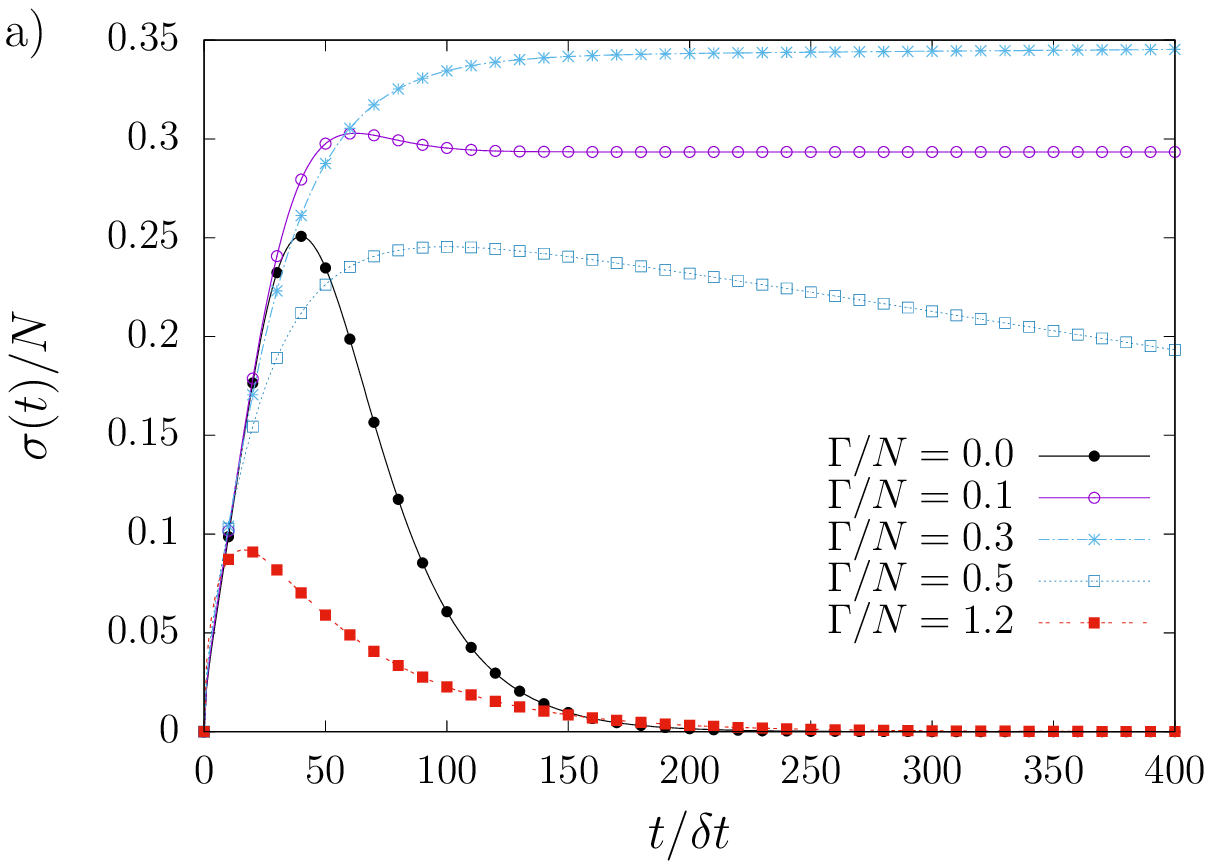}\\
  \includegraphics[width=0.80\columnwidth]{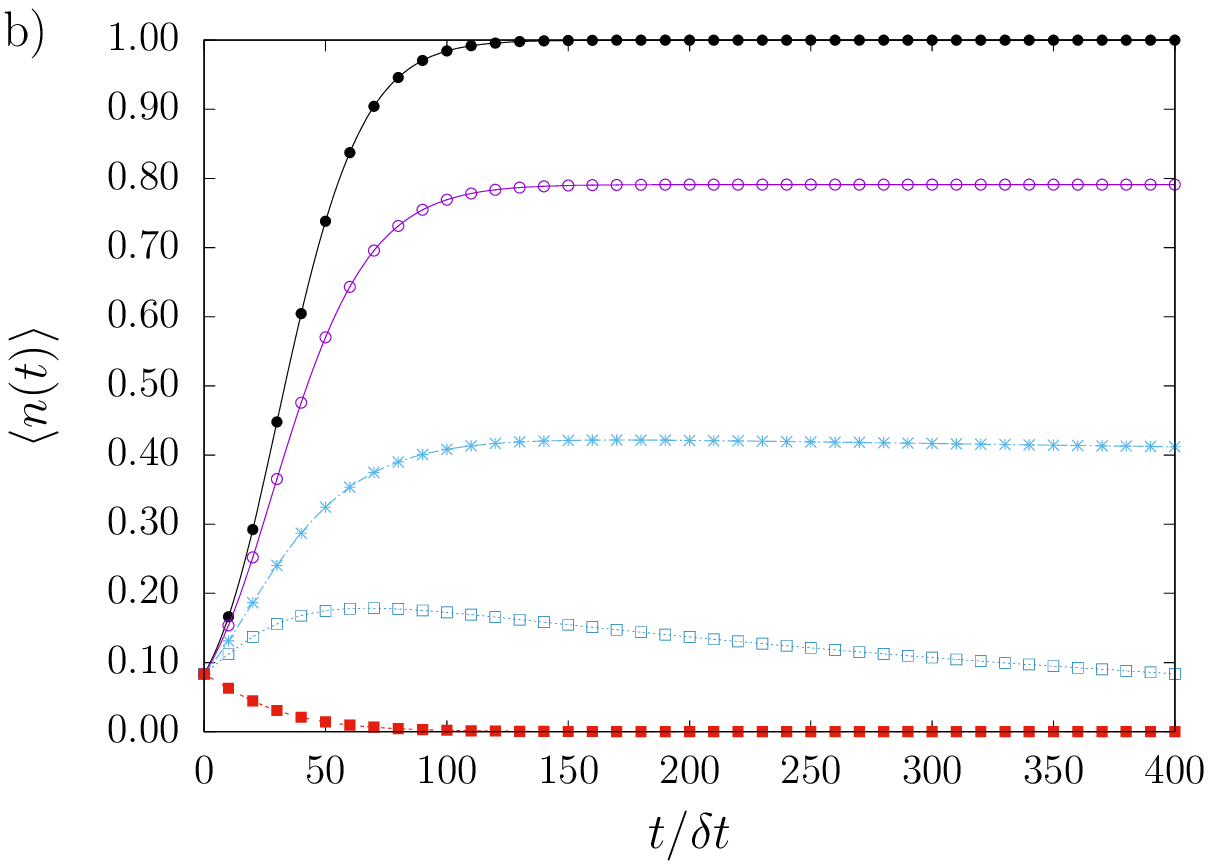}
  \caption{\label{fig:mean} Standard deviation $\sigma(t)$ and mean
    $\langle n(t)\rangle$ for SIS model with $N=12$ in the fully
    connected network. The statistics $\sigma(t)$ and $\langle
    n(t)\rangle$ are shown in a) and b), respectively. Intermediate
    cure/infection rates $\Gamma/N=0.3$ and $\Gamma/N=0.5$ eradicate the
    disease after very large time intervals:  $\sigma(t)$
    exhibits initial rapid growth, develops a maximum at 
    $t'_c\equiv t'_c(\Gamma/N)$ and then decays as power-law.}
\end{figure}

Up to $O(\delta t^2)$, $P_{\mu}(t)$ obeys the following system of
differential equations,
\begin{equation}
\frac{d P_{\mu}}{d t} = -\sum_{\nu} H_{\mu \nu}P_{\nu}(t),
\label{eqdinamica}
\end{equation} 
where $\hat{H}\equiv (\mathbbm{1}-\hat{T})/\delta t$ is the time
generator. For time independent $\hat{H}$, $\lvert 
P(t)\rangle = \exp(-\hat{H} t)\lvert P(0)\rangle$ is the solution of
Eq.~(\ref{eqdinamica}).  The operator $\hat{H}$ governs the dynamics with
eigenvalues $\{ \lambda_{\mu} \}$ and the respective left $\{ \chi_{\mu}\}$ and right
$\{\phi_{\mu}\}$ eigenvectors. The eigenvalues
satisfy $\lambda_{\mu}\geq 0$, vanishing for stationary states
\cite{vanKampen1981}.  
Statistics for observable  $\hat{O}(t)$ are calculated according to $\langle O(t)
\rangle = \sum_{\mu} \langle C_{\mu} \vert \hat{O}\vert C_{\mu}\rangle
P_{\mu}(t)$. Among  the relevant observables in 
disease spreading models, the mean number of infected agents,
$\langle \hat{n}(t)\rangle$, and variance, $\sigma^2(t)$, exemplified in
Fig.~{\ref{fig:mean}}, are often relevant variables. Formally,
they admit eigendecomposition: $\langle n(t)\rangle = 
\sum_{\mu\nu}\gamma_{\mu\nu} \mathe^{-\lambda_{\nu} t}$ and
$\sigma^2(t)=\sum_{\mu\nu}\xi_{\mu\nu}\mathe^{-\lambda_\nu t}-\langle n(t)\rangle^2$, with
$\gamma_{\mu\nu}=\langle C_{\mu}\vert \sum_k\hat{n}_k\vert C_{\mu}\rangle
\langle C_{\mu}\vert \phi_{\nu}\rangle \langle \chi_{\nu}\vert
P(0)\rangle$ and $\xi_{\mu\nu}=\langle C_{\mu}\vert \sum_{k l}\hat{n}_k\hat{n}_l\vert
C_{\mu}\rangle \langle C_{\mu}\vert \phi_{\nu}\rangle \langle
\chi_{\nu}\vert P(0)\rangle$.

\begin{figure}[htbp!]
  \includegraphics[width=0.8\columnwidth]{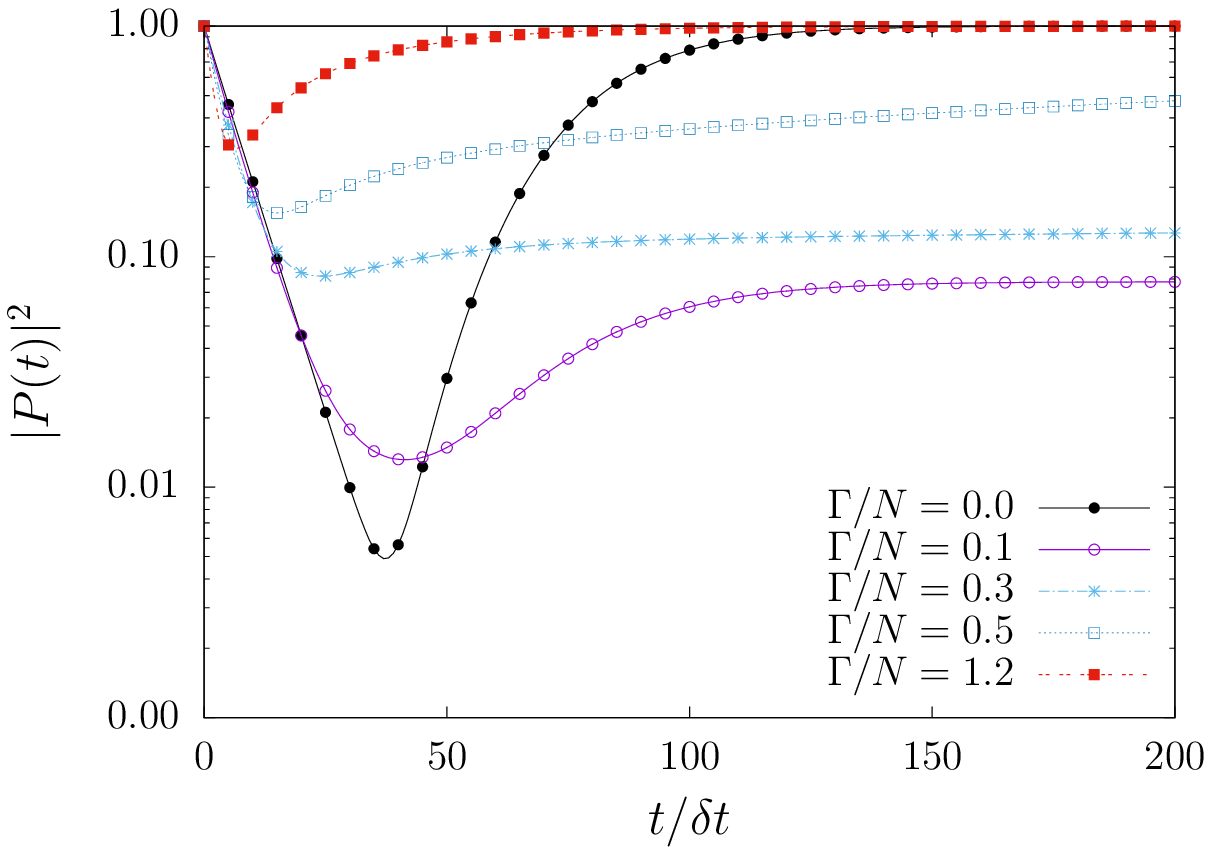}
  \caption{\label{fig:norm} $|P(t)|^2$ in SIS model with $N=12$ and
    $P_1(0)=1$. For each coupling parameter $\Gamma/N$, 
    $| P(t)|^2$ always develops a global minimum followed by constant
    value at stationary state, being unity only for  single state
    configurations. Logarithmic scale is employed
    to emphasize extremal points at $t_c\equiv  t_c(\Gamma/N)$.} 
\end{figure}

\begin{figure}
\includegraphics[width=0.8\columnwidth]{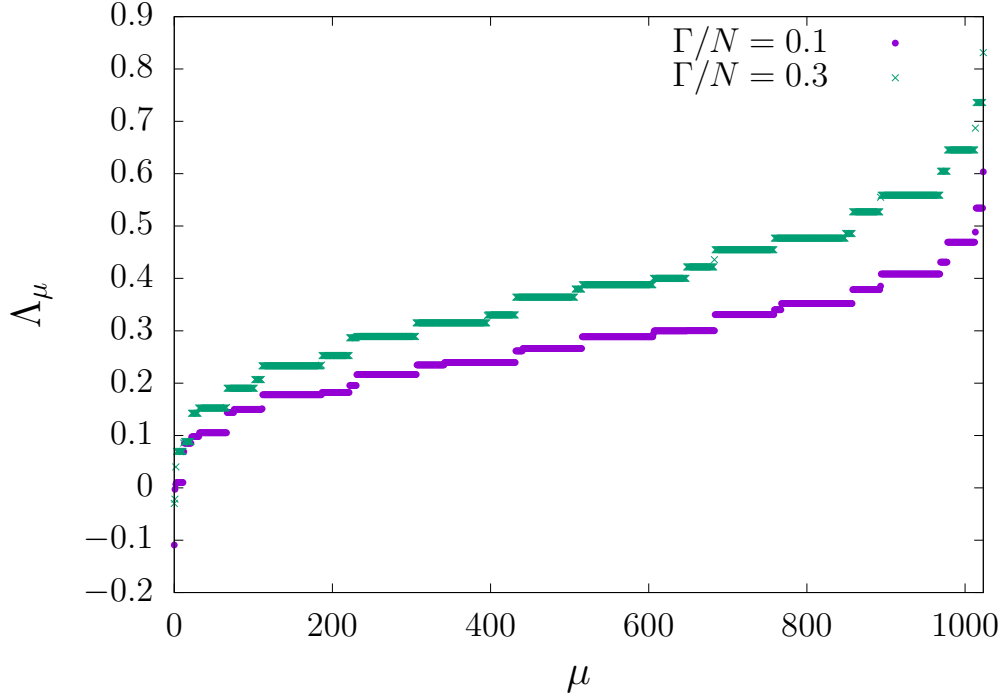}
\caption{\label{fig:eigenvals} Sorted eigenspectra with $N=10$. Each filled circle
  (cross) represents one eigenvalue $\Lambda_{\mu}$ for coupling parameter
  $\Gamma/N=0.1$ ($0.3$).  }
\end{figure}

Although left and right eigenvectors are expected to decompose the
identity, their actual computation is rather cumbersome, doubling the
computational effort and are specially prone to convergence errors. They 
also lack a clear analytical interpretation. Here we consider the
squared norm, $\lvert P(t)\rvert ^2=\sum_{\mu}P_{\mu}^2 $, which remains
invariant under unitary transformations. First, total probability
conservation $\sum_{\mu} P_{\mu}(t) =1$ does not warrant $| P(t) | ^2$
conservation over time. Examples are found in Markov processes that
evolve to unique stationary states since more configurations are
available during the transient. The epidemic models considered in this
study fall into this category, as shown in Fig.~\ref{fig:norm}.
 Second, the
time derivative of $\lvert P(t)\rvert ^2$ is  
obtained from the hermitian operator $\hat{\mathcal{H}}= (1/2)
(\hat{H}+\hat{H}^T)$, 
\begin{equation}
\frac{1}{2}\frac{d}{dt} \lvert P(t)\rvert^2= -
\langle P(t)\rvert \hat{\mathcal{H}} \lvert P(t)\rangle.
\label{eqsqnorm}
\end{equation}
Unlike $\hat{H}$, the operator $\hat{\mathcal{H}}$ has
eigenvalues $\{\Lambda_{\mu}\}$ but the left eigenvectors are computed
from right eigenvectors $\{\psi_{\mu}\}$ by simple Hermitian
conjugation. The trade-off is that $\Lambda_{\mu}$ may assume negative
values, as shown in Fig.~\ref{fig:eigenvals}, and the coefficients
$\langle \psi_{\mu}\vert P(t) \rangle= g_{\mu}(t)$ are complex
numbers.  As such,
the coefficients 
$g_{\mu}$ are not probabilities. Despite this shortcoming, the
coefficients $g_{\mu}$ are
used to evaluate configuration probabilities:
\begin{equation}
P_{\mu}(t)=\sum_{\nu}g_{\nu}(t)\langle C_{\mu}\vert \psi_{\nu}\rangle.
\end{equation}

An important expression is derived from Eq.~(\ref{eqsqnorm}),
\begin{equation}
\sum_{\mu}\left(\frac{1}{2}\frac{d}{d t}+\Lambda_{\mu} \right)|g_{\mu}(t)|^2=0,
\label{eqeig1}
\end{equation}
subjected to the constraint $\sum_{\mu\nu}\langle C_{\mu}
\vert \psi_{\nu} \rangle g_{\nu}(t)=1$.  Now, Eq.~(\ref{eqeig1})
takes a simpler form if $\lvert P(t)\rvert^2$  is constant, which is the
expected outcome whenever the system reaches at least one stationary
state. In such case, Eq.~(\ref{eqeig1}) reads
\begin{equation}
\sum_{\mu}\vert\tilde{g}_{\mu,l}\vert^2\Lambda_{\mu}=0,
\label{stat5}
\end{equation}
where the collection of coefficients $\tilde{g}_{\mu,l}\equiv
\lim_{t\rightarrow   \infty}g_{\mu}(t)$  describes the $l$-th
stationary state. Table~\ref{tab:si_stat} displays Eq.~(\ref{stat5})
non-trivial solution for SI model with $N=3$ and $\beta/N=0.1$. This
simple example is chosen
 since the solution can be evaluated by brute
force and tested against the correct answer. Of course, the trivial
solution $\tilde{g}_{\mu,0}=\delta_{\mu,0}$ and $\Lambda_0=0$ also
satisfy Eq.~(\ref{stat5}). 
\begin{table}
  \caption{\label{tab:si_stat} Stationary state. Non-vanishing
    coefficients $\tilde{g}_{\mu,1}$ in the SI model with $\beta/N=0.1$
    and $N=3$. The set $\{\tilde{g}_{\mu,1}\}$ is obtained solving
    Eq.~(\ref{stat5}). The coefficients are real and  
    $\tilde{g}_{\mu,1}=\langle C_7|  \psi_\mu\rangle$, hence,
    $\tilde{P}_{7}=1$. 
}
\begin{ruledtabular}
  \begin{tabular}{cdd}
    $\mu$ & \Lambda_{\mu} & \tilde{g}_{\mu,1}\\
    3& 0.157199(3)& 0.397770(3)\\
    6& 0.351413(7)&-0.380366(0)\\
    7&-0.108613(0)& 0.834925(5)
  \end{tabular}  
  \end{ruledtabular}
\end{table}

\begin{figure}
  \includegraphics[width=0.8\columnwidth]{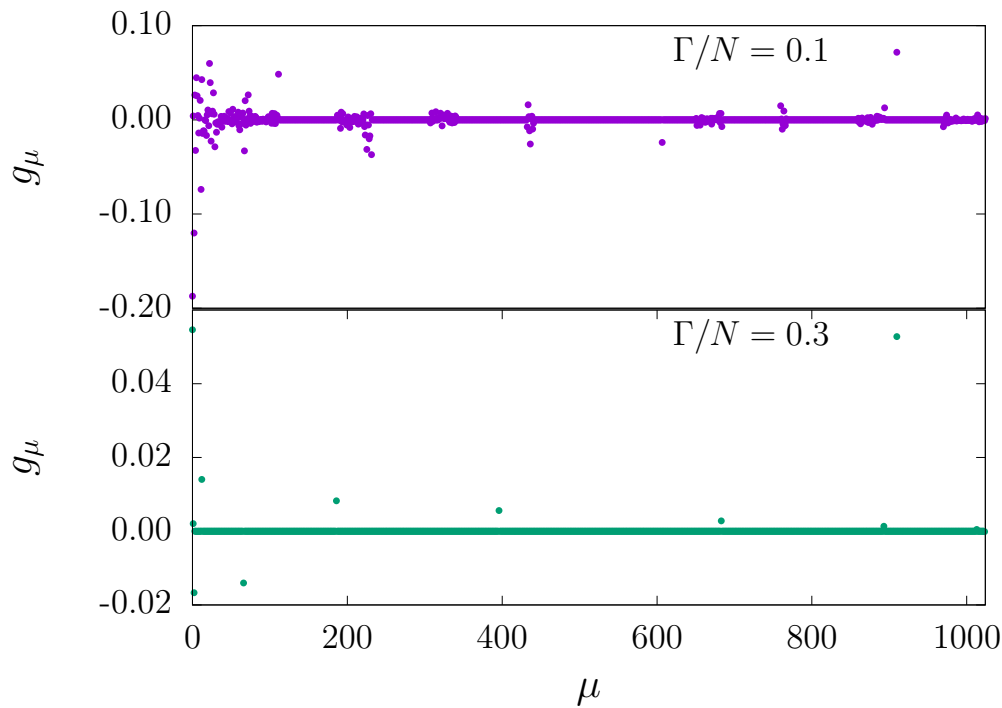}
  \caption{\label{fig:g} Solutions $g_{\mu}(t_c)$ in SIS model with
    $N=10$. Non-trivial solutions to Eq.~(\ref{stat5}) are show for
    $\Gamma/N=0.1, \,0.3$. The coefficient distribution greatly
    differs depending on coupling parameter $\Gamma/N$. 
  }
\end{figure}

In addition to stationarity, $\lvert P(t)\rvert^2$ may also assume
maximal or minimal values at time instants $t_c$, leading again to
Eq.~(\ref{stat5}), the difference being only the evaluation of
coefficients $g_{\mu}(t)$ at $t=t_c$. Numerical examples are shown in Fig.~\ref{fig:g}.
The time instant $t_c$ is
important for dynamics as the extremal condition $\lvert
P(t_c)\rvert^2$ informs us when the disease spreading rate changes its
growth pattern. Accordingly, $t_c$ may also be used to estimate the
maxima for narrow peaked statistics. For instance,  the nonexistence of
cure creates a rapid transient phase in SI model, with all agents
infected as stationary  state. During the transient, the variance
$\langle n^2(t)\rangle - \langle n(t)\rangle^2$ is well described by a
narrow function with peak near $t_c$. The estimation improves as $N$
increases. Therefore, by solving the constrained algebraic
Eq.~(\ref{stat5}), either directly or via functional minimization, one 
also evaluates crucial statistics. 

% Further numerical results using Eq.~(\ref{stat5}) are discussed
% elsewhere.

We note Eqs.~(\ref{eqeig1}) and (\ref{stat5}) introduce a
novel way to tackle stochastic problems: asymmetric operators are
replaced by symmetric operators and the eigenspectra are used to
evaluate stationary states in Eq.~(\ref{stat5}). 
% Each solution
% obtained for Eq.~(\ref{stat5}) corresponds to a unique stationary
% state.
Furthermore, the stationary states obtained in this way
carry the network topological information as the adjacency matrix
determines the eigenvalue distribution. 
In the large $N\gg 1$ regime, the eigenspectrum becomes dense and it is
convenient to analyze Eq.~(\ref{eqeig1}) using the continuous variable
$\Lambda$. For completeness sake, we briefly discuss this regime in
Appendix \ref{sec:app1}. 
Alternatively, one may consider the extremal $|P(t_c)|^2$  and obtain
$t_c$. In turn, $t_c$ may be employed to estimate the time at which
statistics develop maxima, as long as they are narrow peaked functions. 
Since the method is valid for any Markov process, it can be  employed
for more realistic epidemic models.

\section{Perturbation theory}
\label{sec:perturbation}

The eigenvalues $\Lambda_{\mu}$ are crucial to Eq.~(\ref{stat5}) whereas
the eigenvectors $\lvert \psi_{\mu}\rangle$ are required to ensure the
probability conservation constraint. In this section, we consider small
perturbations to the network link distribution and their corresponding
effects on disease spreading in the SIS model.

For the SIS model, the Hermitian time generator is
$\hat{\mathcal{H}}=(\beta/N)[\hat{\mathcal{H}}_0 + \hat{\mathcal{H}}_1]$, with  
\begin{align}
\label{eqsishh0}
  \hat{\mathcal{H}}_0=&\sum_{k j} A_{j k} \left[    ( 1
    -\hat{n}_j) \hat{n}_k - \frac{ \hat{\sigma}_j^+ \hat{n}_k
                        + \hat{n}_k   \hat{\sigma}^-_j }{2} \right]  ,\\ 
  \hat{\mathcal{H}}_1=&\sum_{k j} {\Gamma \delta_{k  j}
                        \left(
                        \hat{n}_k-\hat{\sigma}_k^x
                        \right)}.
\label{eqsishh1}
\end{align}
Here, the adjacency matrix elements $A_{j k}$ are the network
average. The fully connected network is obtained
taking $A_{jk}=(1-\delta_{jk})$ with mean field time generator
$\hat{\mathcal{H}}_{\text{MF}}$. Despite its simplicity, this network 
provides relevant operatorial content. Defining the many-body spin
operators as 
$S^z=\sum_k{\hat{n}_k}-N/2$,
$S^{\pm}=\sum_k{\hat{\sigma}_k^{\pm}}$, and
$S^x=\sum_k{(\hat{\sigma}_k^{+}+\hat{\sigma}_k^{-})/2}$, the resulting
time generator is 
\begin{align}
% \frac{\hat{\mathcal{H}}_{\text{MF}}}{\beta/N}=&\frac{N}{2}\left[\frac{N}{2}-\Gamma
% \right]-\hat{S}^z\left[ \hat{S}^z -\Gamma  \right]+\nonumber\\
% -&\frac{\hat{S}^+\hat{S}^z+\hat{S}^z\hat{S}^-}{2}-\hat{S}^x\left[
%   \frac{N}{2} +\Gamma \right].
\frac{\hat{\mathcal{H}}_{\text{MF}}}{\beta/N}=&\frac{N}{2}\left[\frac{N}{2}-\Gamma
\right]-\hat{S}^z\left[ \hat{S}^z -\Gamma  \right]+\nonumber\\
-&\frac{1}{2}\left\{ \hat{S}^x,\hat{S}^z\right\}-\hat{S}^x\left[
  \frac{N-1}{2} +\Gamma \right].
\label{eqhmeanfield}
\end{align}
The operator $\hat{\mathcal{H}}_{\text{MF}}$ satisfies
$[\hat{\mathcal{H}}_{\text{MF}},\hat{S}^2]=0$, where
$\hat{S}^2=(\hat{S}^z)^2+(1/2)\{\hat{S}^+,\hat{S}^-\}$ is
the Casimir operator. Thus, the eigenvalues $s(s+1)$ are
suitable labels, 
with $s\ge 0$ and $s=N/2,N/2 -1,\ldots$.

An important property is derived from the identification with
 many-body angular momentum operators. The operator 
$\hat{\mathcal{H}}_{\text{MF}}$ in Eq.~(\ref{eqhmeanfield}) prohibits
transitions among different $s$-sectors. This means the
$\hat{\mathcal{H}}_{\text{MF}}$ is block diagonal, each block with dimension
$d=2s+1$. In addition, each block is also tridiagonal in the basis
$\lvert s, m\rangle$ ($m=-s,-s+1,\ldots,s$) as the $\hat{S}^x$ operator
may only increase or decrease $m$ by unity. Therefore, the largest
$s$-sector block has, at most, dimension $d_{\text{max}}=N+1$ and $3N-2$
non-null matrix elements, thus sparsity $O(3/N)$. When both properties
are considered, one realizes the $O(2^{2N}) $ computational problem
has been reduced to $O(N)$. {These properties eliminate one of the
  main obstacles   faced by microscopic agent based epidemic models.}

\begin{figure}
  \includegraphics[width=0.8\columnwidth]{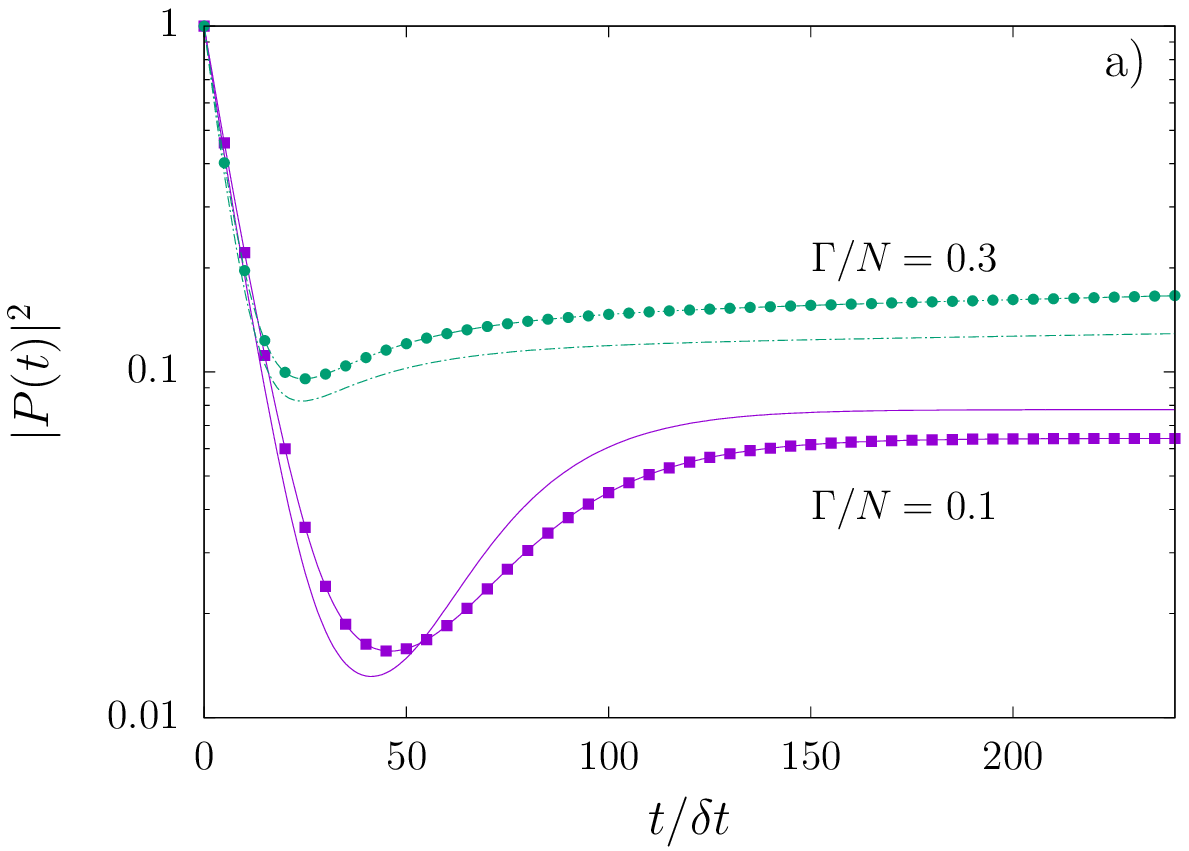}\\
  \includegraphics[width=0.8\columnwidth]{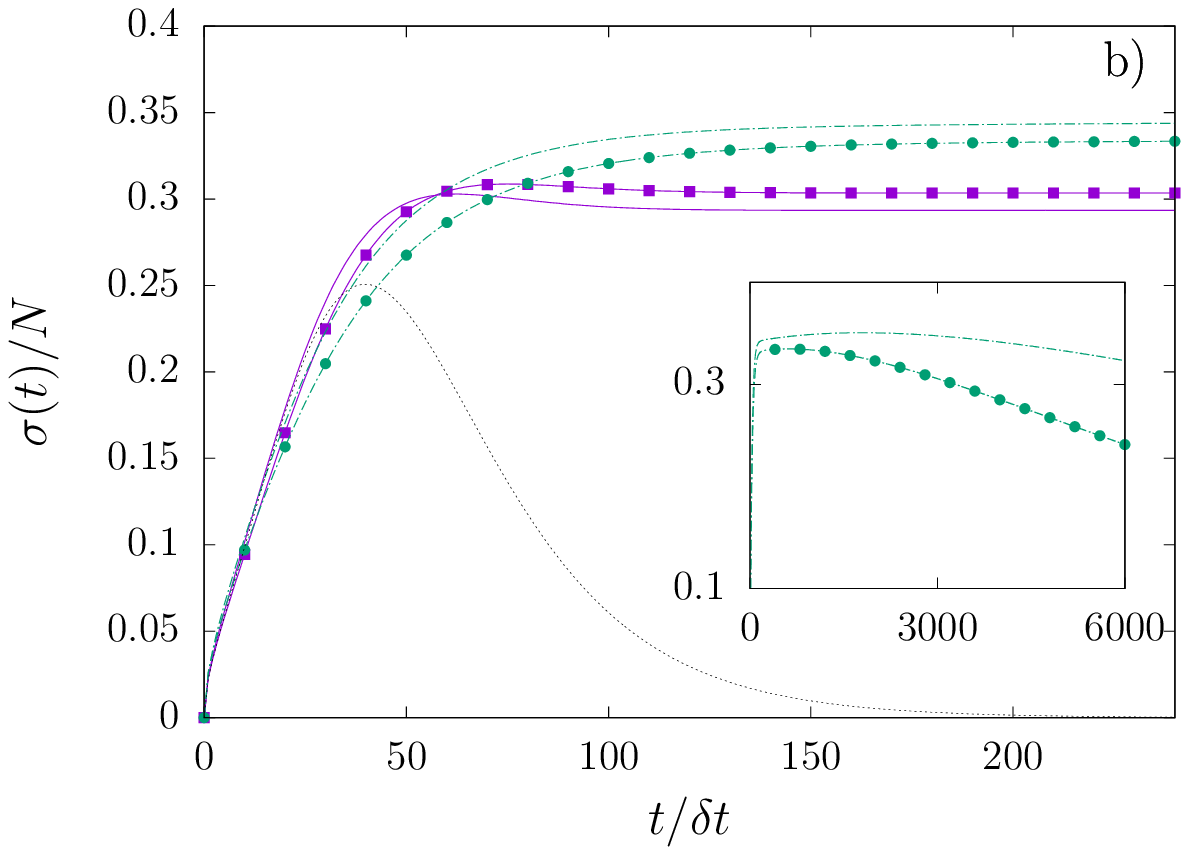}
  \caption{\label{fig:perturbation1} Perturbation in meanfield SIS model and $N=12$.
    In a),  $|P(t)|^2$ is plotted against time steps for SIS coupling
    parameter $\Gamma/N=0.1$ in the meanfield network $\delta p=0$
    (full magenta line) and in the perturbed network $\delta
    p=0.1$ (magenta squares); the corresponding quantities are
    also shown for $\Gamma/N=0.3$ with $\delta p=0$ (dashed green line)
    and $\delta p=0.1$ (green circles). Topological perturbations
    decrease (increase) $|P(t)|^2$ for $\Gamma/N=0.1$ ($0.3$).
    In b), $\sigma(t)$ is plotted against time steps. The dotted line
    displays the expected SI behavior for comparison. The inset shows
    $\sigma(t)$ with $\Gamma/N=0.3$ and $\delta p=0$ (dashed green line)
    and $\delta p=0.1$ ( green circles) using increased time
    range. During transients, small perturbations $\delta p$ may
    produce large modification to the statistics.
}
\end{figure}

For general networks, the main strategy is to use the eigenvectors of
$\hat{S}^2$ and treat any absent link among agents as perturbations.
% The second property concerns the initial state. Since transitions among
% distinct $s$-sectors are forbidden, one may infer the minimum density of
% infected agents by inspecting the initial state. For instance, disease
% eradication is not possible if the initial state has $s < N/2$, since
% the minimum number of infected agents is $n_{\text{min}}=-s + N/2$.
A simple perturbation to the fully connected topology is obtained by
considering a small probability $\delta p\ll 1$ to independently remove
links between vertices, $A_{j k}=(1-\delta_{jk})(1-\delta
p)$. This
procedure is equivalent to transform the underlying network into a random
network \cite{durretPNAS2010}, with connection probability $1-\delta p$.  
Perturbative effects to $|P(t)|^2$ and $\sigma(t)$ are shown in 
Fig.~\ref{fig:perturbation1}. Although both network and topological
perturbations are simple, distinct perturbative effects for
increasing $\Gamma$ are observed.
The perturbative operator is $-\delta p  (\beta/N) \hat{V}$ where
% \begin{align}
%   \hat{V}=&  \left[
%   \left(\frac{N}{2}\right)^2 
%   -(\hat{S}^z)^2 \right] + \nonumber\\
%   -& \left[
%   \left(\frac{N-1}{2}\right)\hat{S}^x +
%   \frac{1}{2}\left\{\hat{S}^z,\hat{S}^x\right\}
%   \right],
%      \label{eqVmean}
% \end{align}
\begin{equation}
  \hat{V}=  
  \left(\frac{N}{2}\right)^2 
  -(\hat{S}^z)^2   
  -  \left(\frac{N-1}{2}\right)\hat{S}^x -
  \frac{1}{2}\left\{\hat{S}^z,\hat{S}^x\right\}
  ,
  \label{eqVmean}
\end{equation}
which is the SI symmetrized time generator and also satisfies
$[\hat{V},\hat{S}^2]=0$.

Concerning stationary states, the first order correction to the eigenvalues, 
$\Lambda_{\mu}^{(1)}=\langle \psi_{\mu} \vert\hat{V} \vert
\psi_{\mu}\rangle$, and eigenvectors,
${g}_{\mu}^{(1)}=\sum_{\nu}^{\prime} \langle \psi_{\mu}\vert\hat{V}
\vert \psi_{\nu}\rangle /(\Lambda_{\nu}-\Lambda_{\mu})$, are obtained
using standard perturbation theory \cite{sakurai1994}. Accordingly,
first order correction to the probability to find the system in
configuration $C_{\mu}$ is 
\begin{equation}
 P_{\mu}^{(1)}=\sum_{\nu}{ \langle   C_{\mu}\vert \psi_{\nu}\rangle
  {g}^{(1)}_{\nu}}.
\label{eqp1}
\end{equation} 
Eq.~(\ref{eqp1}) emphasizes the role played by the Hermitian
operators $\hat{\mathcal{H}}$ to evaluate the effects caused by
topological perturbations:    $P^{(1)}_{\mu}$ and further perturbative
corrections are entirely  evaluated from eigenvectors $\{ \lvert
\psi_{\mu}\rangle \}$ and  eigenvalues $\{ \Lambda_{\mu} \}$, given the
perturbation $\hat{V}$. This rationale suits decision making strategies
as the only concern is the impact of changes to the system topology. The
advantage in our approach lies in avoiding computations of the
asymmetric operator $\hat{H}$ while benefiting from standard
perturbative techniques.

\section{Regular network}

\begin{figure}[ht]
  \begin{tabular}{ccc}
    \includegraphics[width=0.4\columnwidth]{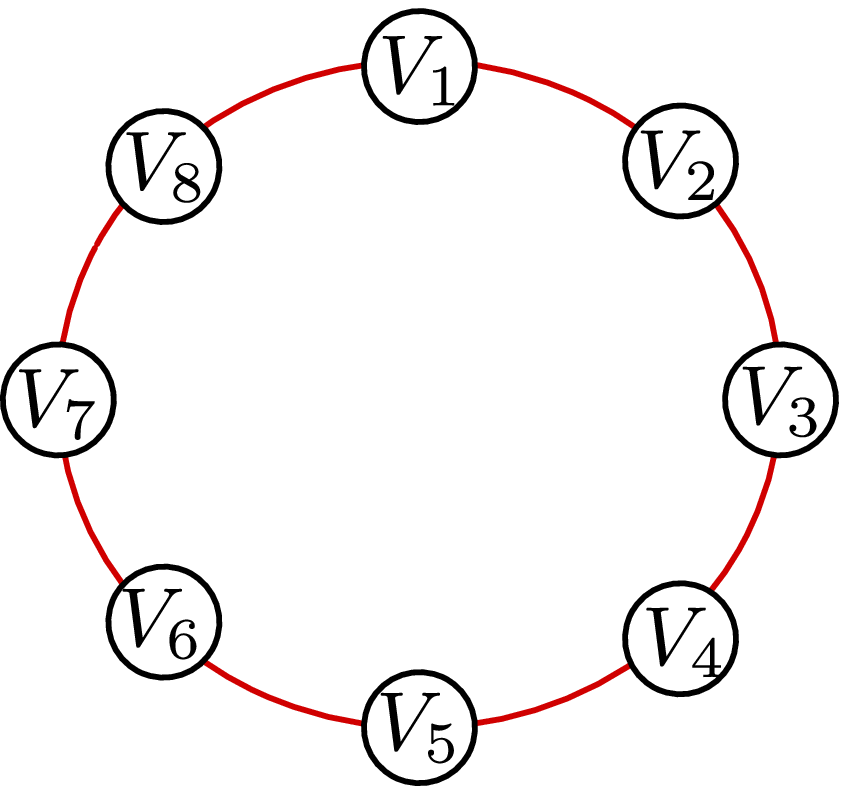} &
    &\includegraphics[width=0.4\columnwidth]{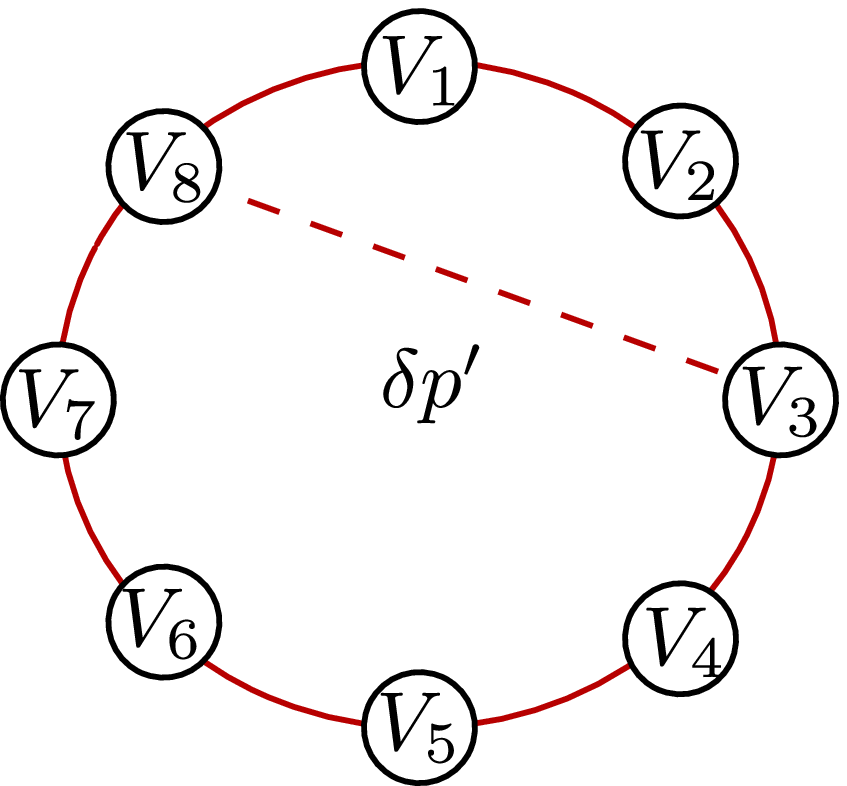}\\
    a) & &b) 
  \end{tabular}
  \caption{One-dimensional periodic lattice with $N=8$ vertices. a) Vertex
  $V_k$ connects with vertex $V_{k+1}$ and $V_{k-1}$ with periodic boundary
  conditions, $V_{N+1}=V_1$ and $V_0=V_N$. Perturbative link addition
  with probability $\delta p = \delta p'/(1+\delta p')$ increases the mean degree
  $d(k)$ by $p(N-2)$, allowing long range disease transmissions. In b),
  the graph shows the regular connections for $V_3$ and an additional
  connection to $V_8$. } 
\label{fig:onedim}
\end{figure}

Fully connected networks are the simplest instances of a larger set known
as regular networks. Other relevant element in the same set is
obtained when the connection patterns among vertices are
periodic. Lattices are their spatial representation and are widely
employed to describe translation invariant systems. Their natural
eigenset contains long and short range modes, allowing analytical tools
to inspect long and short range disease spreading behavior, their
characteristic frequencies and long range  correlations. 
Since perturbative effects are our main concern here, we only
consider a network with single period, or equivalently, a
one-dimensional lattice of size $N$ with periodic boundary condition, as
Fig.~\ref{fig:onedim}a) illustrates. The adjacency matrix is $A_P$ and
the matrix elements are $(A_P)_{k k'}=(\delta_{k, k'+1}+\delta_{k, k'-1})$,
with $V_0=V_N$ and $V_{N+1}=V_1$. 

The perturbative scheme to the network topology adds connections with
probability $\delta p\ll 1$ among vertices not previously connected, as
shown in Fig.~\ref{fig:onedim}b). The perturbation creates shortcuts
throughout the network, favoring rapid disease dispersion, in an attempt
to mimic the relevant aspects found in small-world networks \cite{strogatzNature1998}.  
For a single graph realization, translation invariance breaks and the
important expression  $N^{-1}\sum_k\langle \hat{O}_k\rangle = \langle
\hat{O}_1\rangle$ is no longer valid for a general observable
$\hat{O}_k$. However, for a large ensemble of graphs, the average
transition matrix recovers translation invariance. The reasoning behind
this claim lies in the fact all vertices would have $2+\delta p (N-2)$
neighbors, on average.

Let  $p_{k,k'}=\delta p$ be the probability to create a single link
between $V_{k}$ and $V_{k'}$, including nearest-neighbor
vertices. Clearly, the idea is to emphasize the emergence of translation 
invariance and to interpret the perturbation operator as the meanfield
disease spreading operator $\hat{V}$ in Eq.~(\ref{eqVmean}).  Under this
assumption, the contributions to the adjacency matrix due to
perturbations are $\delta p (1-\delta_{k k'})$. One must be careful to
subtract contributions from links already accounted by
$A_P$, resulting in the symmetric time generator 
$\hat{\mathcal{H}}_{P} =(\beta/N) [ (1-\delta
p)\hat{\mathcal{H}}_0+\hat{\mathcal{H}}_1+\delta p \hat{V} ]$.  
Next, define the effective couplings $\beta'=(1-\delta p)\beta$ and
$\delta p'=\delta p / (1 - \delta p)$, so that
% \begin{align}
%   \frac{\hat{\mathcal{H}}_{P}}{\beta'} &= \frac{1}{\beta}\left(
%                                           \hat{\mathcal{H}}_0+\hat{\mathcal{H}}_1
%                                           \right) \nonumber\\
%   +&\left( \frac{\delta p}{1+\delta p} \right)\left[ \hat{V}-\hat{\mathcal{H}}_1 \right].
% \end{align}
\begin{equation}
  \frac{\hat{\mathcal{H}}_{P}}{\beta'/N} =
  \hat{\mathcal{H}}_0+\hat{\mathcal{H}}_1                                           
  +\delta p'\left( \hat{V}-\hat{\mathcal{H}}_1 \right),
\end{equation}
i.e., the perturbation operator is proportional to $\delta p'$. The
solution for $\delta p'=0$ is obtained using techniques from strongly
correlated systems and spinchains, in momentum space
\cite{liebmattis,nakamuraJPhysA2010}. Moreover, total momentum
$Q=0,1,\ldots,N-1$ is conserved and serves as a label, breaking
$\hat{\mathcal{H}}_{P}$ into $N$ block-diagonal matrices.  For very
large $\delta p$, the network topology moves towards meanfield topology
and favors perturbative analysis using Eq.~(\ref{eqhmeanfield}) as the
unperturbed operator. Therefore, for $\delta p \ll 1$, the perturbative
regime favors periodic eigenvectors whereas for $(1-\delta p)\ll 1 $,
many-body angular momentum eigenvectors are  preferred.

\section{Bethe-Peierls approximation}

In general, perturbations to topology are not required to affect all
vertices in the same manner. For instance, consider a network whose
links are distributed according to a parametric probability density
function $p(\omega)$. If the network undergoes a parameter change  $\omega\rightarrow
\omega+\delta\omega$, one may expect ${A}_{ij}\rightarrow
{A}_{ij}+\delta\omega\, G_{i j}$. The matrix $G$ carries all
modifications experienced by the network under the change. Accordingly,
the symmetric time generator is
$\hat{\mathcal{H}}=(\beta/N)(\hat{\mathcal{H}}_0+\hat{\mathcal{H}}_1+\delta\omega
\hat{V}^{G})$. The perturbation $\hat{V}^G$ is 
\begin{equation}
  \hat{V}^G=\frac{1}{2}\sum_{k j}G_{j k}\left[ 2(1-\hat{n}_j)
    \hat{n}_k
    -\hat{\sigma}_j^+\hat{n}_k-\hat{n}_k\hat{\sigma}_j^-
\right].
\end{equation}
Hermiticity is sufficient to warrant Rayleigh–Schr\"odinger perturbation
theory. Furthermore, Eq.~(\ref{stat5}) requires first order perturbative
corrections $\Lambda_{\mu}^{(1)}$ and $g_{\mu}^{(1)}$ must satisfy
\begin{equation}
  \sum_{\mu}\left[ 
    2\text{Re}(g_{\mu}^* g_{\mu}^{(1)})\Lambda_{\mu}+
    |g_{\mu}|^2\Lambda_{\mu}^{(1)}
\right]=0.
\label{stat6}
\end{equation}
Here, as usual, $\Lambda_{\mu}^{(1)}=\langle \psi_{\mu} \vert\hat{V}
\vert \psi_{\mu}\rangle$ and $g_{\mu}^{(1)}=\sum_{\nu}^{\prime}\lvert \langle
\psi_{\mu}\vert\hat{V} \vert \psi_{\nu}\rangle
\rvert /(\Lambda_{\nu}-\Lambda_{\mu})$.

In addition to perturbative methods, analytical and numerical techniques
from many-body theories are now available to epidemic models. This is
also true for approximations, such as Bethe-Peierls
\cite{betheProcRoySoc1935}. In this approximation,  $\hat{n}_k$ is
replaced by global average $\bar{n}$. Application to the SIS model in an
arbitrary network  produces the effective time generator 
\begin{equation}
\frac{2\hat{\mathcal{H}}'}{\beta/N}= \Gamma
N+\bar{n}\sum_j\kappa_j+\sum_j \Omega_j({\cos\theta_j
    \hat{\sigma}^z_j-\sin\theta_j \hat{\sigma}^x_j}),
\label{effh}
\end{equation}
where $\kappa_j=\sum_k {A}_{k j}$ is the degree of $j$-th vertex,
$\Omega_j=\sqrt{2(\Gamma^2+\bar{n}^2 \kappa_j^2)}$,
$\cos\theta_j = (\Gamma-\bar{n}\kappa_j)/\Omega_j$ and $\sin\theta_j =
(\Gamma+\bar{n}\kappa_j)/\Omega_j$. The effective generator in
Eq.~(\ref{effh}) is diagonalized by rotations around the $y$-axis.

\section{Conclusion}
\label{sec:conclusion}

% Fluctuations are integral part in stochastic processes.
In compartmental approaches to epidemics, the role of fluctuactions is
underestimated 
when the population of infected agents is scarce. Disease
spreading models using agent based models are limited to small
population sizes due to asymmetric time generators and their large
$O(2^{2N})$ dimensions.  Our findings show 
 $\lvert P(t) \rvert^2$ is sufficient to avoid the
mathematical hardships that accompany asymmetric operators.
The squared norm provides a novel way to obtain stationary states and
extremal configurations in general Markov processes, including epidemic
models. Once stationary states are secured, the standard Rayleigh-Schr\"odinger
perturbative technique becomes available to epidemics, making use of
symmetrized operators and their eigenvalues and eigenvectors.
The method paves the way for evaluation of corrections to configuration
probabilities caused by perturbations in complex topologies, where
analytical results are scarce.  
\begin{acknowledgments}
  We are grateful for TJ Arruda comments during manuscript
  preparation. The authors acknowledge Brazilian agencies for
  support. A.S.M. holds grants from CNPq 485155/2013 and 307948/2014-5,
  G.C.C. acknowledges CAPES 067978/2014-01.
\end{acknowledgments}

\appendix

\section{Continuous spectral equation}
\label{sec:app1}

In the large $N\gg 1$ regime, the eigenspectra becomes dense and it is
convenient to analyze Eq.~(\ref{eqeig1}) using the continuous variable
$\Lambda$. Let $\rho(\Lambda)$ be the density of states between
$\Lambda$ and $\Lambda+\delta \Lambda$. In addition, 
consider the real spectral functions $\eta_1(\Lambda, t)$ and
$\eta_2(\Lambda,t)$ so that $g_{\mu}(t)\rightarrow \eta_1(\Lambda,
t)+\imath\, \eta_2 (\Lambda, t)$, with squared norm $\eta^2(\Lambda,t)
\equiv \eta_1^2(\Lambda,t) +\eta_2^2(\Lambda,t) $. 
Since the time evolution of $\lvert P(t)\rvert^2$ is deterministic, it
is convenient to define the functional $S[P]$ over a time interval $t_1-t_0$, 
\begin{equation}
\label{eq:spectral0}
S[P]=\int_{t_0}^{t_1}dt\,\lvert P(t)\rvert ^2,
\end{equation}
with
\begin{equation}
\label{eq:spectral1}
\lvert P(t)\rvert^2 = \int_{-\infty}^{\infty} d\Lambda \rho(\Lambda)
\eta^2(\Lambda,t). 
\end{equation}

Equation~(\ref{eq:spectral0}) suggests the interpretation of $S[P]$ as
the system action. Let the underlying network link distribution be a
continuous function of the real parameter $q$. Next, one considers
virtual variations $\delta q$ to $q$, which produce the
change $\delta\rho(\Lambda,t)$ in the density of states. 
According to Eq.~(\ref{eqeig1}), the continuous variables $\Lambda$ and
real functions $\rho$ and $\eta$ satisfy the following spectral equation:
\begin{equation}
\label{eq:spectral2}
  \int_{-\infty}^{\infty}d\Lambda\, \left(
    \frac{1}{2}\frac{\partial}{\partial t} +\Lambda 
  \right) \rho(\Lambda)  \eta^2(\Lambda,t)=0.
\end{equation}

%\bibliography{bibdatabase}

 %
\end{document}